\documentclass[useAMS,usenatbib]{mn2e}
\usepackage{graphics}
\usepackage{graphicx}
\usepackage{verbatim}

\begin{document}
\title{Imaging sub-milliarcsecond stellar features with intensity interferometry using air Cherenkov telescope arrays}
\author{Paul D. Nu\~nez$^{1}$\thanks{E-mail: pnunez@physics.utah.edu}, Richard Holmes$^{2}$, David Kieda$^{3}$, Janvida Rou$^{4}$, Stephan LeBohec$^{5}$\\
$^{1,3,4,5}$University of Utah, Dept. of Physics \& Astronomy, 115 South 1400 East, Salt Lake City, UT 84112-0830, USA\\
$^{2}$Nutronics Inc., 3357 Chasen Drive, Cameron Park, CA 95682, USA\\
}
\maketitle

\begin{abstract}
Recent proposals have been advanced to apply Imaging Air Cherenkov Telescope (IACT) arrays to Stellar Intensity Interferometry.  Of particular interest is the possibility of model-independent image recovery afforded by the good $(u,v)$-plane coverage of these arrays, as well as recent developments in phase retrieval techniques. The   capabilities of these instruments used as SII receivers have already been explored  for simple stellar objects \citep{mnras}, and here the focus is on reconstructing stellar images with non-uniform radiance distributions. We find that   hot stars ($T>6000^\circ\,\mathrm{K}$) containing hot and/or cool localized regions ($\Delta T\sim 500^\circ\,\mathrm{K}$) as small as $\sim 0.1\,\mathrm{mas}$  can be imaged  at short wavelengths ($\lambda=400\,\mathrm{nm}$).
\end{abstract}


\section{Introduction}

Stellar Intensity Interferometry (SII) has seen a revival due to the extraordinary $(u,v)$ coverage that future air Cherenkov telescope arrays will provide \citep{holder2}. The angular resolution that can be achieved is as fine as $0.06\,\mathrm{mas}$ at the longest baselines ($1.4\,\mathrm{km}$) and the shortest optical wavelengths ($\sim 400\,\mathrm{nm}$) \citep{mnras}. The possible improvement in angular resolution by an order of magnitude and increased sensitivity for hot ($>6000^{\circ}\,\mathrm{K}$) stellar objects \citep{dainis.spie}, has motivated the exploration of SII capabilities to investigate several science topics. These include diameter measurements, stellar rotation, gravity darkening, mass loss and mass transfer (see \citet{dainis.spie} for more details).\\

Intensity interferometry consists in measuring the squared modulus of the complex mutual degree of coherence between detector pairs from the cross-correlation between the light intensity fluctuations. Therefore, only the magnitude of the Fourier transform of the stellar radiance distribution is accessible. This poses a challenge for performing model-independent imaging. However, recent developments in phase recovery techniques from simulated SII data will make model independent imaging possible \citep{mnras, Holmes.spie}. In previous work, we have shown that over-all shapes and dimensions can be reconstructed with sub-percent accuracies by using a Cauchy-Riemann phase reconstruction algorithm \citep{mnras}. The use of this phase reconstruction algorithm along with iterative post-processing routines also allows for further details to be imaged.  \\

This paper is organized as follows: Section 2 outlines the generation of simplified pristine stellar images and the simulation of intensity-interferometry data. Then phase recovery is briefly outlined and image post-processing is presented in Section \ref{cr}. Results are then presented in Section \ref{results} for images with increasing degrees of pristine image complexity.

\section{Pristine stellar images and data simulation}\label{simulations}

Pristine images of disk-like stars are first generated\footnote{The original ``pristine'' image consists of $2048\times2048$
pixels corresponding to $\sim10\,\mathrm{mas}\times10\,\mathrm{mas}$ of angular extension and a
wavelength of $\lambda= 400\,\mathrm{nm}$.}. These images correspond to black-bodies of a specified temperature containing an arbitrary number
of ``star spots'' of variable size, temperature, and location at the surface of the spherical star in this three dimensional model. The simulated 
stellar surface is then projected onto a plane, so that
spots located near the edge of the visible half-sphere appear more elongated than those located near the center. Additionally, limb-darkening is 
included by assuming that the stellar atmosphere has a constant opacity. More details of the simulated images are presented in section \ref{results}.\\

From these pristine images, simulated SII array data are obtained by computing the Fourier transform of the  ``pristine'' image so that the squared modulus of the degree of coherence can be found between 
telescope pairs in a large IACT array \footnote{A preliminary design of the Cherenkov telescope array (CTA) was used \citep{cta, cta_simulation}. See 
Figure 2 in \citep{mnras} for the exact array configuration used in these simulations.}. Gaussian noise is 
then added in the simulations. The signal-to-noise depends on squared modulus of the degree of correlation, the area  
of each of the light receivers, the spectral density (number of photons per unit area per 
unit time per unit frequency), the quantum efficiency, the electronic bandwidth, and the observation time (see \citet{mnras} for more details.). More detailed simulations which include finite detector size, photodetector pulse shape and excess noise are currently under development (Rou, Nu\~nez \& LeBohec, in preparation).\\

The simulations used here do not include the change in baseline due to earth's rotation, and yet exposure times are typically of the order of $10\,\mathrm{hrs}$. These simulations can be regarded as corresponding to the accumulation of a sequence of short ($< 1\,\mathrm{hr}$) observations taken over several ($\sim 10$) nights. It is also possible to save the recorded signal histories in sequences as short as 5 minutes or less, and then this data can be  correlated in post-processing to avoid smearing of information in the $(u,v)$ plane. The simulated stellar images have corresponding temperatures of $\sim 6000^{\circ}\,\mathrm{K}$, and the total exposure time can be significantly reduced if the source has a higher temperature. In fact, most stars for which photon correlations can be detected within $\sim 100\,\mathrm{hrs}$ with a large IACT array, have temperatures greater than $\sim 6000^{\circ}\,\mathrm{K}$ (see section 2 in \citet{mnras} for details).


\section{Cauchy-Riemann phase recovery and post-processing routines}\label{cr}

The Cauchy-Riemann phase recovery algorithm consists in using the theory of analytic functions to relate the magnitude and the phase of the Fourier transform\citep{Holmes}. That is, in 1-dimension, the Cauchy-Riemann equations relate the magnitude and the phase differentials along the real and imaginary directions in the complex plane. For details on the two-dimensional Cauchy-Riemann phase recovery algorithm see \citep{mnras}, section 5. The 
resulting reconstructed image with this estimated phase is sometimes not ideal, and so is taken as a first guess
for the iterative algorithms that are now described.\\

\begin{figure}
\begin{center}
\includegraphics[scale=0.5]{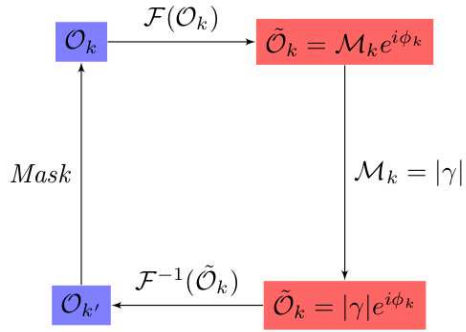}
\end{center}
\caption{\label{gs_routine}Schematic of the Gerchberg-Saxton error reduction algorithm.}
\end{figure}

The Gerchberg-Saxton algorithm \citep{gs}, also known as the error-reduction algorithm, is an iterative procedure. Starting from a reasonable guess of the image, 
the algorithm consists in going back and forth between image and Fourier space,  each time imposing general restrictions in each 
domain. Figure \ref{gs_routine} describes the Gerchberg-Saxton algorithm. Starting from an image $\mathcal{O}_k$, the first step consists in 
taking the Fourier transform to obtain something of the form $\mathcal{M}_ke^{i\phi_k}$. Next, Fourier constraints can then be applied, i.e. the magnitude is 
replaced by that given by the data, and the phase of the Fourier transform is maintained. Next, the inverse Fourier transform is calculated and constraints can 
be imposed in image space. The constraints in image space can be very general. The image constraint that we impose comprises applying a 
\emph{mask}, so that only pixels within a certain region are allowed to have positive non-zero values. For the images presented below, the mask is a circular region whose radius is typically found by measuring the radius of the first guess obtained from the Cauchy-Riemann approach. In all reconstructions where the Gerchberg-Saxton is used, we perform 50 iterations, and found that more iterations typically do not produce significant changes in the reconstruction. \\ 

Another post-processing application that has been utilized is \small{MIRA} (Multi-aperture image reconstruction algorithm) \citep{Thiebaut}. \small{MIRA}
has become a standard tool for image reconstruction in amplitude (Michelson) interferometry. \small{MIRA} is an iterative procedure which slightly modifies image
pixel values so as to maximize the agreement with the data. In the image reconstruction process, additional constraints such as smoothness or compactness can be applied simultaneously, but this is something that we have not yet experimented with, i.e. the regularization parameter is set to zero for all reconstructions presented here. In the results presented below, the number of iterations is set by the default stopping criterion of the optimizer. The \small{MIRA} software only uses existing data in the $(u,v)$ plane, as opposed to using the fit of an analytic function to the data as is done in the Cauchy-Riemann and Gerchberg-Saxton routines. This results in removing artifacts in the reconstruction that can be caused by the fit of an analytic function to the data. \citet{renard} have recently investigated imaging capabilities with \small{MIRA} in the context of amplitude interferometry, and the results shown below are presented in such a way so that they can be compared with amplitude interferometry capabilities.

\section{Results}\label{results}

Pristine images are generated with varying complexity. We first generate pristine images of stars with limb-darkened atmospheres and investigate
the reconstruction capabilities. Then we 
introduce a localized bright or dark feature, and finally increase the number of features and explore some of the parameter space, 
i.e. spot size, location, etc. 

\subsection{Limb-darkening}\label{limb_darkening}

\begin{figure}
  \includegraphics[scale=0.5, angle=-90]{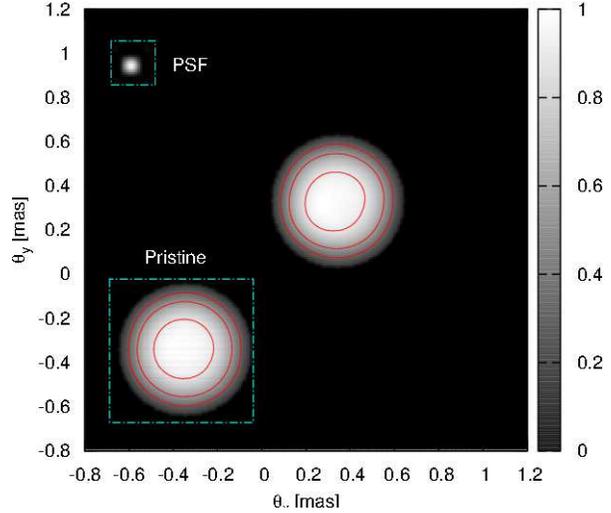}
  \caption{\label{limb_example} Image reconstruction of a star with limb darkening parameter $\alpha=5$, apparent visual magnitude $m_v=3$ 
and $10\,\mathrm{hrs}$ of observation time. The pristine starting image from which intensity interferometric data were simulated 
is shown in the bottom left corner with the same contour lines. The Cauchy-Riemann phase reconstruction was performed to produce a raw image, and then the Gerchberg-Saxton
routine was implemented to produce the post-processed image shown.}
\end{figure}

\begin{figure}
  \includegraphics[scale=0.3, angle=-90]{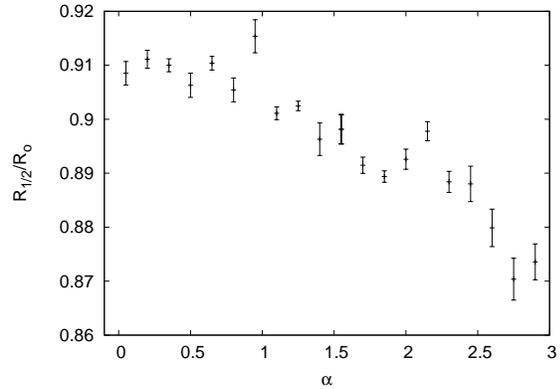}
  \vspace{0.5cm}
  \caption{\label{r_vs_alpha} For each value of $\alpha$, SII data were 
simulated corresponding to stars with apparent visual magnitude $m_v=3$, $10\,\mathrm{hrs}$ of observation time and $\lambda=400\,\mathrm{nm}$. For 
each image reconstruction, a 1-dimensional profile is found by calculating the average intensity at each radial position. The 
 half height $R_{1/2}$ (angular radius at 
half maximum) is found. For each value of $\alpha$, the process was repeated 10 times, and error bars are the standard deviation of the distribution 
of $R_{1/2}(\alpha)$. The ratio $R_{1/2}/R_o<1$ since the image reconstruction is at best a convolution with the PSF of the array. }
\end{figure}

Image reconstruction is actually not necessary for the study of limb-darkening, which can be studied with the knowledge of the squared degree of coherence 
only. A direct analysis of the data, with no image reconstruction, is likely to yield better results than the ones presented below. However, it is instructive to first see this effect in reconstructed images before adding stellar features to the simulated pristine images. Limb darkening can be 
approximately modeled with a single parameter $\alpha$ as $I(\phi)/I_0=(\cos{\phi})^\alpha$ \citep{limb}, 
where $\phi=0$ refers to light being emitted from the center
of the stellar disk, and directed radially to the observer. The values of $\alpha$ depend on the wavelength and  can be found by assuming 
hydrostatic equilibrium. At a wavelength of $400\,\mathrm{nm}$, $\alpha\approx 0.7$ \citep{limb_wavelength} for Sun-type stars, and deviations from this value may be indicative of stellar mass loss. An example of the reconstruction of a limb darkened star with $\alpha=5$ is shown in Figure \ref{limb_example}; such a large value is chosen so that the effect is clearly visible in a two dimensional image with linear scale. More realistic values are considered below. To obtain Figure
\ref{limb_example}, a first estimate of the phase was obtained from the Cauchy-Riemann algorithm and used to generate a raw image. Then the 
Gerchberg-Saxton post-processing loop (Figure \ref{gs_routine}) was performed several (50) times.  In  Figure \ref{r_vs_alpha} the ratio of the average radius at half maximum $R_{1/2}$ and the nominal radius of $R_o$, is shown as a function of the limb darkening parameter $\alpha$. Here data were simulated corresponding to stars with apparent visual magnitude $m_v=3$, temperature $T=6000^\circ\,\mathrm{K}$, radii of $R_o=0.3\,\mathrm{mas}$, $10\,\mathrm{hrs}$ of observation time and $\lambda=400\,\mathrm{nm}$. The ratio $R_{1/2}/R_o$ is less than 1, even in the absense of noise since the reconstruction is at best a convolution\footnote{A ``perfect'' reconstruction gives $R_{1/2}/R_o=0.96$ for $\alpha=0$. The extra discrepancy is due to a small hot-spot in the reconstruction. $R_{1/2}<R_o$ since the reconstruction is normalized to the highest pixel value. } with the array point-spread-function (PSF). \\

From Figure \ref{r_vs_alpha} we can see  that we are sensitive to changes of order unity in the limb-darkening parameter $\alpha$. Even though this analysis
is not sensitive to small changes in $\alpha$, stars experiencing high mass loss rates are likely to have high values of $\alpha$.  If we fit 
a uniform disk function to a limb-darkened reconstruction, the fit yields a smaller radius. For example, in the case of $\alpha=2.0$, 
a uniform disk fit yields an angular radius that is smaller by 7\% (still larger than 
the sub-percent uncertainties found in radius measurements \citep{mnras}). A 
real example is the case of the star \emph{Deneb}, where
the difference between the extracted uniform disk diameter ($\theta_{UD}=2.40\pm0.06\,\mathrm{mas}$) and the limb-darkened 
diameter is $0.1\,\mathrm{mas}$ \citep{deneb}, and its measured mass loss rate is $10^{-7}M_{\odot}\mathrm{yr}^{-1}$.

\subsection{Stars with single features }\label{single_spots}

\begin{figure}
  \includegraphics[scale=0.5, angle=-90]{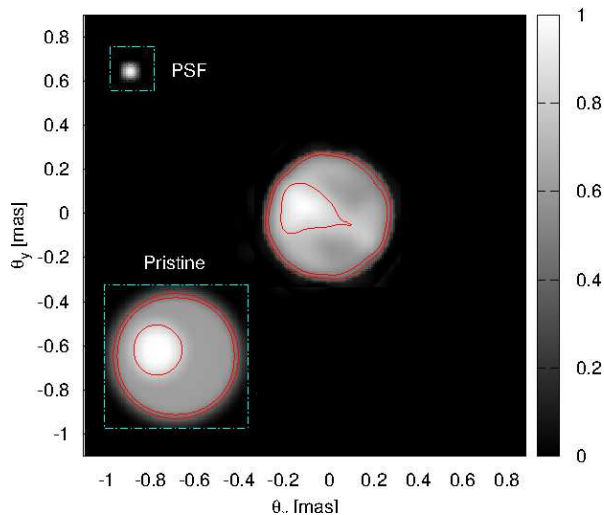}
  \caption{\label{bright_spot}Reconstructed bright spot. This simulated reconstruction corresponds to a star of $m_v=3$, $10\,\mathrm{hrs}$ 
of observation time, $T=6000\,\mathrm{K}$, and spot temperature of $T_{spot}=6500^\circ\,\mathrm{K}$.}
\end{figure}

\begin{figure}
  \includegraphics[scale=0.5, angle=-90]{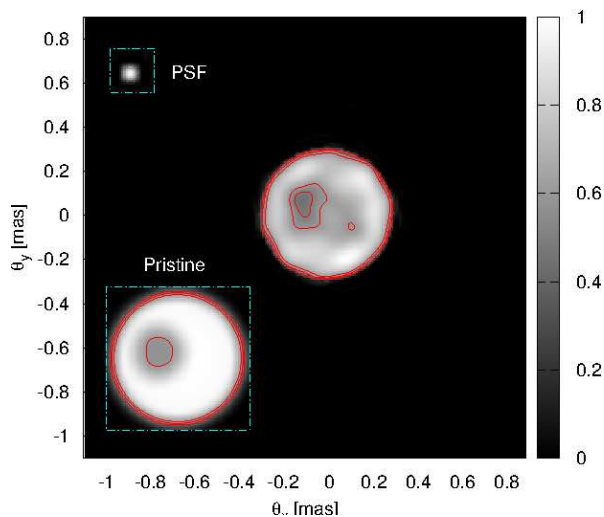}
  \caption{\label{dark_spot} Reconstructed dark spot. This simulated reconstruction corresponds to a star of $m_v=3$, $10\,\mathrm{hrs}$ 
of observation time, $T=6000\,\mathrm{K}$, and spot temperature of $T_{spot}=5500^\circ\,\mathrm{K}$.}
\end{figure}

Stars were simulated as black bodies with a localized feature of 
a higher or lower temperature as described in section \ref{simulations}. In the simulated images, the effect 
of limb darkening is included as described in the previous section. Here the full  reconstruction analysis was used, which consists 
in first recovering a raw image from the Cauchy-Riemann algorithm. The raw image is 
then used as a starting point
for several iterations of the Gerchberg-Saxton loop (see Figure \ref{gs_routine}), and finally the output of the Gerchberg-Saxton algorithm 
is the starting image for the \small{MIRA} algorithm. Examples can be seen in 
Figures \ref{bright_spot} and \ref{dark_spot}, corresponding to the post-processed reconstructions of bright stars
of $m_v=3$, $10\,\mathrm{hrs}$ of observation time and a temperature $T=6000^\circ\,\mathrm{K}$. In Figure \ref{bright_spot} the 
temperature of the spot is $T_{spot}=6500^{\circ}\,\mathrm{K}$, and in Figure \ref{dark_spot} the temperature of the spot is 
$T_{spot}=5500^{\circ}\,\mathrm{K}$. \\

\begin{figure}
  \includegraphics[scale=0.33, angle=-90]{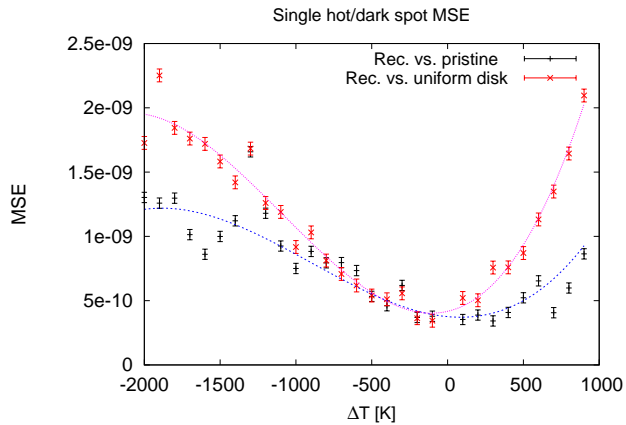}
  \vspace{0.5cm}
  \caption{\label{spot_correlations} The bottom curve and data points correspond to the MSE between reconstructed images and their corresponding
   pristine images containing spots of temperature $T+\Delta T$, where $T=6000^{\circ}\,\mathrm{K}$ ($m_v=3$, $10\,\mathrm{hrs}$ of observation time). The 
   top curve and data points 
   are the MSE comparing the reconstructed image and a uniform disk of the same size as the pristine image. To estimate the uncertainties, we 
  performed several reconstructions and found the statistical standard deviation of the MSE.}
\end{figure}

We can estimate the smallest temperature contrast that can be detected by varying the parameters in the model producing the pristine image. The performance 
in terms of temperature contrast obviously depends on several variables such as the size, location and shape of the spot. To quantify  the smallest detectable spot temperature contrast, we calculate the Mean Square Error (MSE)\footnote{For two images $A_{i,j}$ and $B_{i,j}$ of dimension $N\times N$,  $\mathrm{MSE}=\mathrm{Min}_{k,l}\left\{N^{-2} \sum_{i,j}^{N, N}(A_{i,j}-B_{i+k, j+l})^2\right\}$. Images should be normalized, i.e. $\sum_{i,j}^{N,N}A_{i,j}=\sum_{i,j}^{N,N}B_{i,j}=1$.} comparing the reconstructed image and the pristine image convolved with the array PSF. The ratio of the angular size of the image  and the region over which the MSE is calculated ($0.6\,\mathrm{mas}/1.76\,\mathrm{mas}$) is comparable to the same ratio in the work of \citet{renard} ($34\,\mathrm{mas}/100\,\mathrm{mas}$), so that the MSEs presented below are directly comparable to those obtained by \citet{renard}. The
MSE is still difficult to interpret by itself, so it is compared  with the MSE between the reconstruction and a simulated star with 
no spots. By comparing these
two values we can have an idea of the confidence level for reconstructing spots with different temperatures. This comparison is shown in 
Figure \ref{spot_correlations}, where the bottom curve corresponds to the MSE as a function of spot temperature $6000^\circ\,\mathrm{K}+\Delta T$  
between the reconstructed images and the pristine images, and
the top curve is the MSE comparing the reconstructed images and a spotless disk of the same size as the pristine image. A total of 26 stars 
were simulated, and the uncertainty in the MSE was estimated by performing several (10) 
reconstructions for one particular case ($\Delta T=500^\circ\,\mathrm{K}$). From the Figure
it can be seen that spots are accurately imaged  when $\Delta T<- 700^\circ \, \mathrm{K}$ or 
$\Delta T> 200^\circ \, \mathrm{K}$ approximately.  For a black body of spectral density $B(T)$, a temperature difference  
$\Delta T<- 700^\circ \, \mathrm{K}$ corresponds to a flux ratio
$B(T+\Delta T)/B(T)<0.45$, and a temperature difference  $\Delta T> 200^\circ \, \mathrm{K}$ corresponds to flux ratios 
$B(T+\Delta T)/B(T)>1.2$. This asymmetry can be partly understood in terms of the brightness ratio between black bodies 
$B(T+\Delta T)/B(T)$, whose rate of change is higher when $\Delta T>0$ than when $\Delta T<0$. This however does not fully account for the asymmetry
between cool and hot spots. Most of the asymmetry is due to the fact that all the simulated stars in Figure \ref{spot_correlations} have the same
integrated brightness, and the radiance per solid angle is larger for a star containing a bright spot than for an annular region in a star containing
a dark spot. The same analysis can be performed by simulating stars with different integrated brightness, but the estimate becomes unnecesarily 
cumbersome and implies knowledge that we would not have access to prior to performing an image reconstruction\footnote{For example, we would need to have information
on  the radiance per solid angle in an annular region in a star containing a dark spot.}. We should also not forget that this is an estimate, and in a more precise calculation we would need to consider additional variables such as spot size, position, etc. \\

\begin{figure}
  \includegraphics[scale=0.45]{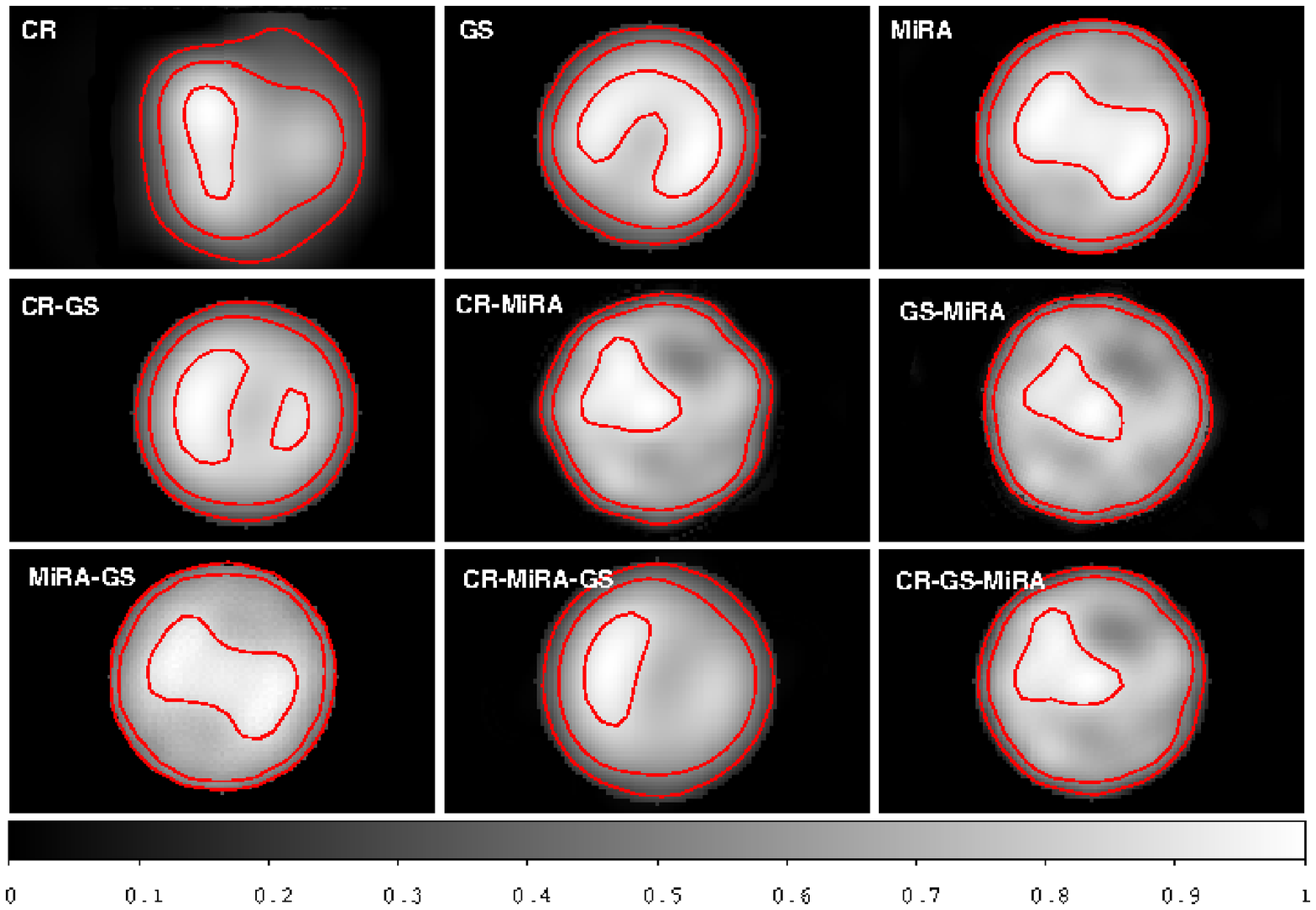}
  \caption{\label{sup1}Image reconstructions for different algorithm combinations. The pristine corresponds to that of Figure \ref{bright_spot}. }
\end{figure}

\begin{figure}
  \vspace{3.2cm}
    \includegraphics[scale=0.45]{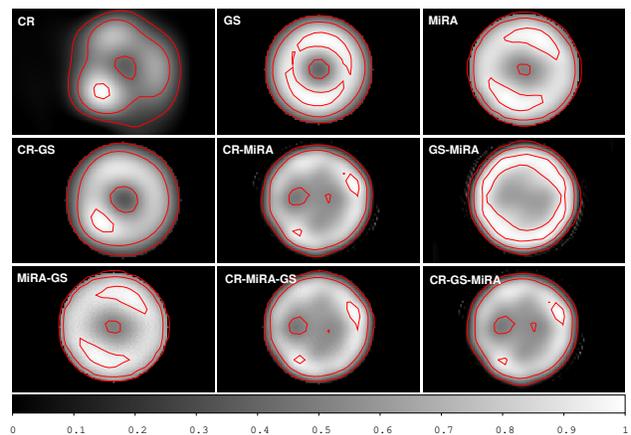}
    \caption{\label{sup2}Image reconstructions for different algorithm combinations. The pristine corresponds to that of Figure \ref{dark_spot}. }
\end{figure}

To test whether the full chain of algorithms is needed to produce Figures \ref{bright_spot} and \ref{dark_spot},  and to investigate algorithm performance, we reconstruct Figs. \ref{bright_spot} and \ref{dark_spot} with different algorithms and combinations of algorithms. Then we calculate the MSE of the reconstructions with the pristine image (convolved with the PSF), and also compare it to the MSE with a uniform disk. The MSE is found for reconstructions using combinations of Cauchy-Riemann\footnote{It only makes sense to use the Cauchy-Riemann algorithm first, since this is not an iterative algorithm relying on a first guess.} (CR), Gerchberg-Saxton (GS), and \small{MIRA}. Whe only \small{MIRA} or Gerchberg-Saxton are used directly (not preceded by Cauchy-Riemann), the initial guess of the image is a uniform disk.  The results are shown in Table \ref{table1}. The resulting reconstruction for each of the algorithm combinations is shown in Figures \ref{sup1} and \ref{sup2}. The single most effective algorithm for these reconstructions is \small{MIRA}, and the lowest MSE is obtained by using Cauchy Riemann, followed by Gerchberg-Saxton and \small{MIRA}. According to table \ref{table1}, the GS algorithm by itself does better than the CR algorithm by itself. This is because the pristine images used here have a high degree of symmetry.  This preferentially aids the GS algorithm. Also the GS algorithm does better on these cases because the support constraint, given by the size of the star, is easy to compute. When \small{MIRA} or the Gerchberg-Saxton algorithms are used directly, the reconstructed image is usually symmetric. The Cauchy-Riemann stage provides sensitivity to asymmetries such as with the bright or dark spot displayed in Figures \ref{bright_spot} and \ref{dark_spot}. The role of Gerchberg-Saxton is more to improve the phase reconstruction. \small{MIRA} plays the important role of removing artifacts, caused for example by the data fitting in the Cauchy-Riemann phase, and  improving overall definition. Even though non-symmetric images can be reconstructed, the final product still is somewhat more symmetric than the pristine image for reasons that are still under investigation. When the MSE with the pristine image is compared to the MSE with a uniform disk, we can again see that the bright spot (Figure \ref{bright_spot}) is more easily detected than the dark spot (Figure \ref{dark_spot}). 


\begin{table}
\begin{center}
\begin{tabular}{||l |l l||} \hline\hline
Algorithm & MSE (Fig. 4) &\\
          & Pristine & UD  \\\hline
CR &   $1.5\times 10^{-8}$  &$1.4\times 10^{-8}$ \\
GS &   $1.1\times 10^{-9}$  &$1.6\times 10^{-9}$ \\
\small{MIRA} &   $4.8\times 10^{-10}$  &$6.7\times 10^{-10}$\\
CR $\rightarrow$ GS &   $1.7\times 10^{-8}$  &$1.7\times 10^{-8}$ \\
CR $\rightarrow$ \small{MIRA} &   $4.4\times 10^{-10}$  &$6.9\times 10^{-10}$\\
GS $\rightarrow$ \small{MIRA} &   $5.4\times 10^{-10}$  &$6.5\times 10^{-10}$\\
\small{MIRA} $\rightarrow$ GS &   $6.0\times 10^{-10}$  &$8.4\times 10^{-10}$\\
CR $\rightarrow$ \small{MIRA} $\rightarrow$ GS &   $5.2\times 10^{-10}$ & $8.6\times 10^{-10}$ \\  
CR $\rightarrow$ GS $\rightarrow$ \small{MIRA} &   $4.7\times 10^{-10}$  &$7.6\times 10^{-10}$ \\
\hline\hline
\end{tabular}

\begin{tabular}{||l |l l|l l||} \hline\hline
Algorithm & MSE (Fig. 5)&\\
          & Pristine & UD  \\\hline
CR &    $1.2\times 10^{-8}$  &$1.3\times 10^{-8}$ \\
GS &    $1.3\times 10^{-9}$ &$1.2\times 10^{-9}$  \\
\small{MIRA}  &  $6.0\times 10^{-10}$  &$4.1\times 10^{-10}$\\
CR $\rightarrow$ GS  &  $1.6\times 10^{-8}$  &$1.7\times 10^{-8}$\\
CR $\rightarrow$ \small{MIRA} &   $5.5\times 10^{-10}$ & $5.5\times 10^{-10}$\\
GS $\rightarrow$ \small{MIRA} &   $6.6\times 10^{-10}$ & $4.4\times 10^{-10}$\\
\small{MIRA} $\rightarrow$ GS  &   $6.1\times 10^{-10}$ & $4.1\times 10^{-10}$\\
CR $\rightarrow$ \small{MIRA} $\rightarrow$ GS  &   $7.0\times 10^{-10}$ & $6.9\times 10^{-10}$\\  
CR $\rightarrow$ GS $\rightarrow$ \small{MIRA}  &   $4.8\times 10^{-10}$ & $6.3\times 10^{-10}$\\
\hline\hline
\end{tabular}

\end{center}
\caption{\label{table1} Mean square errors for different combinations of reconstruction algorithms starting from simulated data corresponding to the pristine image of Figs. \ref{bright_spot} (top table) and \ref{dark_spot} (bottom). The algorithms are CR (Cauchy-Riemann), GS (Gerchberg-Saxton) and \small{MIRA}. The MSE between the reconstructed image and the pristine image (convolved with the PSF of the array) is shown in the second column and is compared to the MSE comparing the reconstructed image an a uniform disk in the third column. The uncertainty in the MSE is $\Delta \mathrm{MSE}=4\times 10^{-11}$.}
\end{table}

A better estimate of the smallest temperature difference that can be imaged requires an exhaustive exploration of parameter space, but temperature 
differences of less than a few hundred Kelvin do not seem to be possible to image when the same brightness, temperature, exposure time,  
angular diameter, spot size and spot position as above are used. Results are likely to improve for hotter stars than those simulated above since signal-to-noise is higher and also the brightness contrast is higher for the same relative temperature differences ($\Delta T/T$). Another question related to imaging single spots is that of finding the smallest spot that can be reconstructed. In previous work \citep{mnras}, we show that the smallest possible spot that can be reconstructed is given by the PSF of the IACT array used in the simulations, namely $0.06\,\mathrm{mas}$.

\subsection{Multiple features}
\begin{figure}
  \includegraphics[scale=0.5, angle=-90]{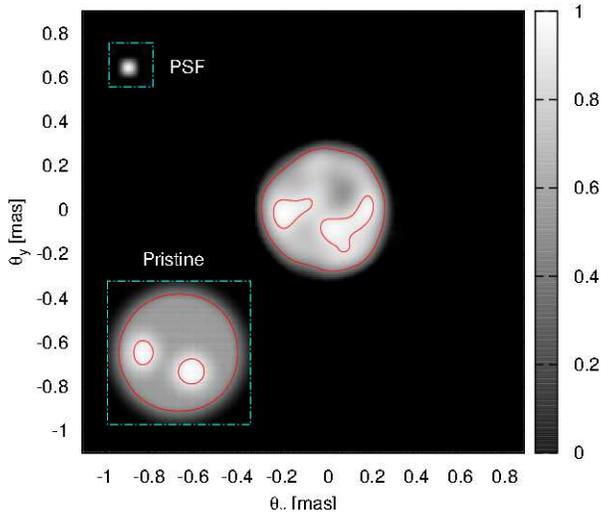}
  \caption{\label{double_spot} Reconstructed star with two hot spots. The pristine image has a temperature of $6000^\circ\,\mathrm{K}$, and each hot spot has a temperature of $6500^\circ\,\mathrm{K}$. The simulated data corresponds to $m_v=3$ and $T=10\,\mathrm{hrs}$. }
\end{figure}

\begin{figure}
  \includegraphics[scale=0.5, angle=-90]{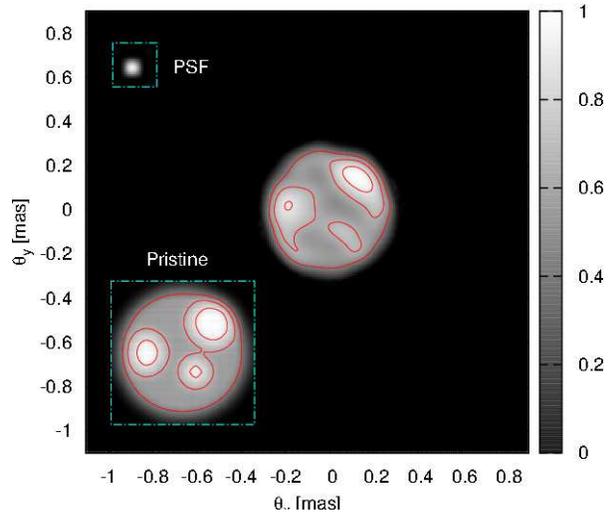}
  \caption{\label{triple_spot} Reconstructed star with three hot spots. The pristine image has a temperature of $6000^\circ\,\mathrm{K}$, and the spots have temperatures of $6500^\circ\,\mathrm{K}$ (top right spot and left spot) and $6800^\circ\,\mathrm{K}$ (lower spot), The simulated data corresponds to $m_v=3$ and $T=10\,\mathrm{hrs}$. }
\end{figure}

As a natural extension to the simulations presented above, data were simulated corresponding to stars with two or more recognizable features. In Figures \ref{double_spot} and \ref{triple_spot},  reconstructions of stars containing several hot spots are shown. The brightness and exposure time are the same as those used to simulate single-spot stars ($m_v=3, T=10\,\mathrm{hrs}$). A detailed investigation of reconstruction of two-spot stars was not performed, but the general behavior is similar to that presented in the section \ref{single_spots}. The reconstructions improve significantly when the pristine image has a higher degree of symmetry, e.g. when both spots lie along a line that splits the star in two. This is expected since phase reconstruction is not really necessary for centro-symmetric images. For this reason, we tested reconstructions with non-symmetric pristine images. Even though the shape of the spots is usually not well reconstructed, the approximate position and size are reasonably accurate.\\

The reconstruction and identification of features degrades as the number of features in the pristine image is increased. A common characteristic of reconstructing stars with several features, is that the larger features in the pristine image are better reconstructed. This is more so in the case of stars containing darker regions. Nevertheless, information of positions, sizes and relative brightness of star spots can still be extracted. In Figure \ref{triple_spot} a reconstruction of a star containing three hot spots of different sizes and relative brightness is shown. This simulated reconstruction corresponds to the same brightness and exposure parameters as those of Figure \ref{double_spot}.

\section{Conclusions}

The capabilities of future IACT arrays for reconstructing complex images were demonstrated via simulations. Stars were simulated as black bodies with temperatures of $6000^\circ\,\mathrm{K}$ with localized hot or cool regions, then data were simulated as would be obtained with an IACT array of $\sim 100$ telescopes separated by up to $1.4\,\mathrm{Km}$. A first raw image obtained with the Cauchy-Riemann phase reconstruction algorithm was performed, followed by post-processing with the Gerchberg-Saxton and \small{MIRA} routines. Post-processing significantly improves the image reconstruction, but the post-processing routines by themselves are usually not sufficient for performing reconstructions, especially when the pristine image is not centro-symmetric. Features with sizes of $\sim 0.05\,\mathrm{mas}$ and temperature differences of a few hundred $^\circ\mathrm{K}$ were reconstructed. Some statistical tests were performed to quantify the confidence with which these features were reconstructed. The resulting mean square errors of the simulated reconstructions are comparable to simulated results obtained by \citep{renard} in the context of amplitude interferometry, but in a different angular resolution regime.\\

Results suggest that a variety of physical phenomena can be imaged. For example, a hot star that is losing mass radiatively can have localized bright and
dark regions corresponding to a higher or lower local mass loss rate \citep{clumps}. The origin of these localized regions is unknown, but is likely related to 
magnetic activity. An order of magnitude estimate\footnote{This can be done within the CAK \citep{cak} formalism for radiatively driven mass loss.} shows that stars with a local radiative flux that is roughly twice as high as the rest of the star, such as those shown in the Figures above, undergo a localized mass loss rate that is three times as high as the rest of the star. A related effect that is within imaging reach is the von-Zeipel effect, in which the local radiative flux is proportional to the surface gravity, so that rotationally distorted stars have brighter poles.  The analysis of 
 featured oblate star observations would be very similar to the one done here. From previous results obtained in \citep{mnras}, 
oblateness ratios (ratio of semi-major and
semi-minor axis) of $10\%-20\%$ are within reach. These in turn correspond to flux ratios of the order of $20\%-40\%$, which have 
also been simulated and reconstructed herein. A more comprehensive study on 
the imagability of these physical phenomena will be the subject of future research. It will be interesting to see how well radiative-transfer models can be tested with the proposed approach.

\end{document}